\documentclass[twocolumn,preprintnumbers,amsmath,amssymb]{revtex4}
\usepackage{amsmath,amsfonts,latexsym,amssymb,graphics,epsfig,subfigure,color,makeidx}
\usepackage{xcolor}
\usepackage{multirow}

\begin{document}\large

\title{Generalization of regular solutions of Einstein's gravity equations and Maxwell's equations for point-like charge
\footnote{\large{Translated by Nikolai Korchagin, a physicist from JINR}}
}

\author{Yi-Shi Duan}
\affiliation{\large Moscow State University}


\begin{abstract} \large
\begin{center}
          Soviet Physics JETP 1954 Vol. 27(6): 756-758\\
\end{center}
{}
\end{abstract}

\maketitle


In the Born [1], Infeld [2], Bopp [3], Podolsky [4] and Dirac [5] theories the electron mass is finite (or zero), but gravity effects have not
been considered. Shirokov [6] and Fisher [7] showed that in studying of origin of elemental particle masses we can not neglect these effects.
Gravity field plays an essential role in interpretation of particle mass. It has been shown [8, 9], that for point-like charge the general
solution of gravity and electromagnetic (EM) equations that contains arbitrary function $\mu(r)$, can in special case of this function,
describe gravity and EM fields that do not have singularities in full range of $r$ from $0$ to $\infty$.

Here we are trying to find more general solution.

General solution of gravity and EM equations for point-like charge leads to [9]:
\begin{eqnarray}
  ds^2 &=& - e^\lambda dr^2 - e^\mu r^2 (d\theta^2 + \sin^2\theta d\phi^2)+e^\nu dt^2, \nonumber \\
  x_1 &=& r,~~x_2=\theta, ~~x_3 = \phi,~~x_t =t,
\end{eqnarray}
where $ds$ is the relativistic interval in the spherical-like coordinates,
$0\leq r \leq +\infty,~~0 \leq \theta \leq \pi,~~ 0 \leq \phi \leq 2 \pi$,
and $\lambda$, $\mu$ and $\nu$ are functions of $r$:
\begin{eqnarray}
g_{11} &=& -e^\lambda,~~~~~~~~~~~~~~~g_{22} = -e^\mu r^2,~~ \nonumber\\
g_{33} &=& -e^\mu r^2 \sin^2\theta, ~~~~~g_{44} = e^\nu,\nonumber\\
g_{ik} &=& 0 ~~\text{for} ~~ i\neq k,  \\
g ~&=& -r^4 \sin^2\theta e^{\lambda+\nu +2\mu},\nonumber
\end{eqnarray}
with
\begin{eqnarray}
e^\lambda &=& \left(1+\dfrac {\mu'r}{2}\right)^2 e^{-\nu+\mu}, \nonumber\\
e^\nu &=& 1 - \dfrac {2m}{r} e^{-\mu/2} +\dfrac {4\pi \varepsilon^2}{r^2} e^{-\mu},\\
T^1_1 &=& - T^2_2 = - T^3 _3 = T^4_4 = \dfrac {1}{2} \dfrac{\varepsilon^2}{r^4} e^{-2\mu}.
\end{eqnarray}
Here the $\varepsilon$ is the particle charge in Heaviside units and $T^i_k$ is the EM field energy-momentum tensor.

For calculation of the proper mass they usually use [10]
\begin{eqnarray*}
  m &=& \int \big(\sqrt{-g}T^4_4+t^4_4\big)d\tau \\
    &=& \int \sqrt{-g} \big(T^4_4 - T^1_1 -T^2_2 -T^3_3 \big)d\tau.
\end{eqnarray*}
Here $t^i_k$ is the energy-momentum pseudotensor for gravity field.

For EM field $T^i_i=0$. Therefore
\begin{equation}
  m = 2 \int \sqrt{-g} T^4_4 d\tau.
\end{equation}

Let's set
\begin{equation}
  \mu = -2 \ln f(r)~~~(\text{or} ~~f(r) = e^{-\mu/2}),
\end{equation}
where $f(r)$ is an arbitrary function without singularities in the region $0 \leq r \leq + \infty$, and satisfies the following conditions:
\begin{eqnarray}
~[f(r)]_{r=0} &=&0, \nonumber \\
~[f'(r)]_{r=0} &=&\alpha,~~~\alpha \neq 0,  \\
~[f''(r)]_{r=0} &=& \beta, ~~~\beta \neq 0. \nonumber
\end{eqnarray}

It is possible to show that the functions $\mu(r)=2 \ln (1+2m/r)$ and $\mu(r)=-2 \ln (1-e^{-kr})$ chosen by Einstain [8] and Shirokov [9] satisfy these conditions.
Substituting Eq. (6) into Eqs. (2)-(4), we get
\begin{eqnarray}
e^\nu ~&=& 1-\dfrac{2m}{r} f(r)+\dfrac{4\pi\varepsilon^2}{r^2} f^2(r),\nonumber \\
e^{\nu+\lambda} &=& \dfrac {[f(r)-rf'(r)]^2}{f^4(r)};
\end{eqnarray}
\begin{eqnarray}
g^{-1}_{11} &=& -e^{-\lambda} \nonumber \\
  &=& - \dfrac {f^4(r)}{[f(r)-rf'(r)]^2} \nonumber \\
   && \times\left[1-\dfrac{2m}{r} f(r)+\dfrac{4\pi\varepsilon}{r^2} f^2(r) \right],
\nonumber \\
g_{22} &=& -e^\mu r^2  = -r^2/f^2(r),
\nonumber \\
g_{33} &=& -e^\mu r^2 \sin^2\theta = -r^2 \sin^2 \theta /f^2(r),
\nonumber \\
g_{44} &=& e^\nu =1-\dfrac{2m}{r} f(r) +\dfrac{4\pi\varepsilon}{r^2} f^2(r);
\end{eqnarray}
\begin{eqnarray}
\sqrt{-g} &=& \dfrac {r^2 \sin\theta |f(r)-rf'(r)|}{f^4(r)}, \nonumber\\
T^4_4 &=& \dfrac 1 2 \dfrac {\varepsilon^2 f^4(r)}{r^4}.
\end{eqnarray}

With the help of Eq. (7) it is easy to get
\begin{eqnarray}
~[g^{-1}_{11}]_{r=0} &=& -(4\alpha^4/\beta^2) \left[1-2m\alpha+4\pi\varepsilon^2\alpha^2 \right],
\nonumber \\
~[g_{22}]_{r=0} &=& -1/\alpha^2,
\nonumber \\
~[g_{33}]_{r=0} &=& -\sin^2 \theta/\alpha^2,
\nonumber \\
~[g_{44}]_{r=0} &=& 1-2m\alpha +4 \pi \varepsilon^2 \alpha^2.
\end{eqnarray}
Equations (11) show that the metric is regular at $r=0$.

Substituting Eq.~(10) into Eq.~(5), we get the particle mass
\begin{eqnarray}
  m &=& 2 \int \sqrt{-g}T^4_4 d\tau \nonumber\\
    &=& - 4 \pi \varepsilon^2 \int_0^\infty d [\dfrac {f(r)}{r}] = 4 \pi \varepsilon^2 \alpha,
\end{eqnarray}
so, the mass is finite.

If we set
\begin{equation*}
  \alpha = 1/r_0,
\end{equation*}
then from Eq. (12) we get
\begin{equation}
r_0 = 4 \pi \varepsilon^2/m = e^2/mc^2,
\end{equation}
where $e$ is the particle charge in the absolute units.
Equation (13) shows that $r_0$ plays role of the classical electron radius.

Note that according to Eqs.~(4), (5) and (12), the whole mass is sum of gravitational and EM field masses.

Finally, I would like to thank Prof. Shirokov for useful discussions.

\end{document}